\documentclass[fleqn,usenatbib]{mnras}
\usepackage{longtable}
\usepackage{graphicx}
\usepackage{amsmath,amssymb}
\usepackage{color}
\usepackage{units}
\usepackage{epstopdf}
\usepackage{hyperref}
\usepackage{multirow}
\usepackage{url}
\usepackage{subfigure}
\usepackage{rotating}
\usepackage{enumitem}\setlist[description]{font=\textendash\enskip\scshape\bfseries}
\usepackage{booktabs}
\usepackage{lineno}

\usepackage{newtxtext,newtxmath}

\usepackage[T1]{fontenc}

\DeclareRobustCommand{\VAN}[3]{#2}
\let\VANthebibliography\thebibliography
\def\thebibliography{\DeclareRobustCommand{\VAN}[3]{##3}\VANthebibliography}

\usepackage{graphicx}	
\usepackage{amsmath}	
\usepackage{amssymb}	

\title[Serendipitous Detections of Kilonovae]{Optimizing Serendipitous Detections of Kilonovae: Cadence and Filter Selection}

\author[M. Almualla et al.]{
Mouza Almualla,$^{1}$\thanks{E-mail: mouzaalmuallaa@gmail.com}
Shreya Anand,$^{2}$
Michael W. Coughlin,$^{3}$
Tim Dietrich,$^{4}$ \newauthor
Nidhal Guessoum,$^{1}$
Ana Sagu{\'e}s Carracedo,$^{5}$
Tom{\'a}s Ahumada,$^{6}$
Igor Andreoni,$^{2}$ \newauthor
Sarah Antier,$^{7}$
Eric C. Bellm,$^{8}$
Mattia Bulla,$^{9,10}$
and Leo P. Singer$^{11,12}$
\\
$^{1}$Department of Physics, American University of Sharjah, PO Box 26666, Sharjah, UAE\\
$^{2}$Division of Physics, Mathematics, and Astronomy, California Institute of Technology, Pasadena, CA 91125, USA\\
$^{3}$School of Physics and Astronomy, University of Minnesota, Minneapolis, Minnesota 55455, USA\\
$^{4}$Institute of Physics and Astronomy, University of Potsdam, Karl-Liebknecht-Str. 24/25, 14476, Potsdam, Germany\\
$^{5}$The Oskar Klein Centre, Department of Physics, Stockholm University, AlbaNova, SE-106 91 Stockholm, Sweden\\
$^{6}$Department of Astronomy, University of Maryland, College Park, MD 20742, USA\\
$^{7}$Universit\'e de Paris, CNRS, Astroparticule et Cosmologie, F-75013 Paris, France\\
$^{8}$DIRAC Institute, Department of Astronomy, University of Washington, 3910 15th Avenue NE, Seattle, WA 98195, USA\\
$^{9}$Nordita, KTH Royal Institute of Technology and Stockholm University, Roslagstullsbacken 23, SE-106 91 Stockholm, Sweden\\
$^{10}$The Oskar Klein Centre, Department of Astronomy, Stockholm University, AlbaNova, SE-106 91 Stockholm, Sweden\\
$^{11}$Astrophysics Science Division, NASA Goddard Space Flight Center, MC 661, Greenbelt, MD 20771, USA\\
$^{12}$Joint Space-Science Institute, University of Maryland, College Park, MD 20742, USA\\
}

\date{Accepted XXX. Received YYY; in original form ZZZ}

\pubyear{2021}

\begin{document}
\label{firstpage}
\pagerange{\pageref{firstpage}--\pageref{lastpage}}
\maketitle

\begin{abstract}
The rise of multi-messenger astronomy has brought with it the need to exploit all available data streams and learn more about the astrophysical objects that fall within its breadth. One possible avenue is the search for serendipitous optical/near-infrared counterparts of gamma-ray bursts (GRBs) and gravitational-wave (GW) signals, known as \textit{kilonovae}. With surveys such as the Zwicky Transient Facility (ZTF), which observes the sky with a cadence of $\sim$\,three days, the existing counterpart locations are likely to be observed; however, due to the significant amount of sky to explore, it is difficult to search for these fast-evolving candidates. 
Thus, it is beneficial to optimize the survey cadence for realtime kilonova identification and  enable further photometric and spectroscopic observations.
We explore how the cadence of wide field-of-view surveys like ZTF can be improved to facilitate such identifications. We show that with improved observational choices, e.g., the adoption of three epochs per night on a $\sim$\,nightly basis, and the prioritization of redder photometric bands, detection efficiencies improve by about a factor of two relative to the nominal cadence. We also provide realistic hypothetical constraints on the kilonova rate as a form of comparison between strategies, assuming that no kilonovae are detected throughout the long-term execution of the respective observing plan. These results demonstrate how an optimal use of ZTF increases the likelihood of kilonova discovery independent of GWs or GRBs, thereby allowing for a sensitive search with less interruption of its nominal cadence through Target of Opportunity programs. 
\end{abstract}




\section{Introduction}

Large field-of-view all-sky surveys will play a central role in the future of time-domain astronomy. Facilities with survey cadences and fields-of-view (FOVs) that will enable such endeavors include the Panoramic Survey Telescope and Rapid Response System (Pan-STARRS; \citealt{MoKa2012}), the Asteroid Terrestrial-impact Last Alert System (ATLAS; \citealt{ToDe2018}), the Dark Energy Camera (DECam; \citealt{FlDi2015}), the Zwicky Transient Facility (ZTF; \citealt{Bellm2018,Graham2018,DeSm2018,MaLa2018}), the enhanced Public ESO Spectroscopic Survey for Transient Objects (ePESSTO; \citealt{SmVa2015}), and the Global Rapid Advanced Network Devoted to Multi-messenger Addicts (GRANDMA; \citealt{AgAl2020, AnAg2020}) and in the near future, BlackGEM \citep{BlGr2015}, Gravitational-wave Optical Transient Observer (GOTO; \citealt{GoCu2020}),  and the Vera C.~Rubin Observatory's Legacy Survey of Space and Time (LSST; \citealt{Ivezic2019}). The access to recent reference images from these surveys renders the discovery of new transients a routine affair. These facilities employ a variety of follow-up telescopes (e.g. \citealt{HoJo2004,BlNe2018,Wilson2003,Coughlin2018}), international consortia of astronomers (e.g. \citealt{AnAg2020,GoCu2020,LuPa2019,KaAn2020}), image difference pipelines (e.g. \citealt{ZaOf2016,hotpants}), and machine learning-based techniques (e.g. \citealt{MuNa2019,Ishida:2018uqu}) to efficiently follow-up the myriad of transients that are found regularly.

These wide FOV facilities have had significant success due to serendipitous discoveries of interesting transients, but also play a crucial role in amplifying the returns in the era of multi-messenger astrophysics.
Although there have been previous examples of serendipitous detections of short gamma-ray burst (SGRB) afterglows from the intermediate Palomar Transient Factory (iPTF) \citep{CeUr2015}, ATLAS \citep{StTo2017}, and ZTF \citep{HoPe2020,KaAn2020}, the use of space-based gamma-ray
observatories to provide localizations is more common-place.
SGRB localizations from the \textit{Fermi} Gamma-Ray Burst Monitor (GBM; \citealt{MeLi2009}) can span $\approx 100-1000$\,deg$^{2}$, making follow-up challenging even for facilities like ZTF, and almost impossible for small FOV telescopes (i.e., those with FOV $\lessapprox$ 1\,deg$^{2}$). However, the \textit{Swift} mission \citep{GeCh2004}, another discovery engine for GRBs, localizes them to areas of a few arcmin\,$^2$ \citep{SaBa2011} with its 1.4 steradian-wide Burst Alert Telescope (BAT) \citep{BaBa2005}, X-ray Telescope (XRT) \citep{BuHi2005}, and UV/Optical Telescope (UVOT) \citep{RoKe2005}. On the other hand, Fermi GBM detects SGRBs at a rate of $\approx$1 week$^{-1}$, which is four times the rate of Swift \citep{SaBa2011, vonKienlin2020}. 



In addition to \textit{Swift} and \textit{Fermi}, there are other instruments actively producing transient alerts with relatively coarse localizations. In particular, these include the Advanced LIGO \citep{aLIGO} and Advanced Virgo \citep{adVirgo} gravitational-wave (GW) detector network, and IceCube \citep{AaAc2017}, which detects neutrino events. These localization regions vary from tens to many thousands of square degrees. Generally, because of their scientific importance, these events are followed up as part of Target of Opportunity programs by survey telescopes (e.g. \citealt{AnAg2020,GoCu2020,LuPa2019,KaAn2020}).
Due to the volume of alerts and limited telescope time, most survey telescopes that perform Target-of-Opportunity observations must be selective about which events they choose to follow up. In the case of ZTF, triggering criteria can include the visibility of the skymap, confidence of the event being astrophysical, and trigger latency \citep{CoAh2019b}.



Compact binary coalescences involving a neutron star, which produce SGRBs, also have a broadly isotropic electromagnetic signature known as a \emph{kilonova} (or \emph{macronova}) \citep{LaSc1974,LiPa1998,MeMa2010,Ro2015,KaMe2017}; see~\cite{Metzger:2019zeh} for a recent review and further references. This kilonova is driven by the radioactive decay of r-process elements in highly neutron rich, unbound matter that can heat the ejecta and power a thermal ultraviolet/optical/near-infrared transient. As an exemplary case, after the detection of GW170817~\citep{AbEA2017b}, both a SGRB counterpart \citep{2017ApJ...848L..21A,2017ApJ...848L..25H,2017Sci...358.1579H,2017ApJ...848L..20M,2017Natur.551...71T} and a kilonova counterpart, AT2017gfo \citep{ChBe2017,2017Sci...358.1556C,CoBe2017,2017Sci...358.1570D,2017Sci...358.1565E,KaNa2017,KiFo2017,LiGo2017,2017ApJ...848L..32M,NiBe2017,2017Sci...358.1574S,2017Natur.551...67P,SmCh2017,2017PASJ...69..101U}, were discovered.

GW170817 has inspired dedicated searches for serendipitous kilonovae (i.e., kilonovae discovered independently of GWs or GRBs) from wide FOV surveys, e.g. Pan-STARRS \citep{McSm2020} and ZTF \citep{andreoni2020constraining}. These searches, although as yet unsuccessful in detecting strong candidates, are empirically constraining the rates of kilonovae.
Serendipitous detections of kilonovae on their own will also enable constraints on the neutron star equation of state \citep{BaJu2017, MaMe2017, CoDi2018b, CoDi2018, CoDi2019b, AnEe2018, MoWe2018,RaPe2018,AbAb2018,Lai2019}, the Hubble constant \citep{CoDi2019,CoAn2020,2017Natur.551...85A,HoNa2018}, and r-process nucleosynthesis \citep{ChBe2017,2017Sci...358.1556C, CoBe2017,PiDa2017,SmCh2017,WaHa2019,KaKa2019}.

Recently, following the systematic search for serendipitous detections of kilonovae during the first 23 months of ZTF described in \citet{andreoni2020constraining}, a new pipeline for ZTF Realtime Search and Triggering (ZTFReST) has been launched in order to identify kilonova-like transients and rapidly trigger photometric follow-up. This pipeline makes use of techniques such as forced photometry and stacking in order to calculate lightcurve evolution rates that can help distinguish between red, fast-evolving kilonova- and afterglow-like candidates and other kinds of kilonova impostors.  Our work complements these significant improvements on kilonova candidate detection pipelines by investigating alternative survey \emph{strategies} that could more efficiently yield kilonovae and multi-messenger counterparts through serendipitous observations. 

In addition to past studies of the detection feasibility of kilonovae for different model parameters and survey facilities \citep[e.g.,][]{RoFe2017,Scolnic_2017}, and explorations of Target-of-Opportunity versus serendipitous searches for kilonovae \citep{Cowperthwaite_2019}, there have also been previous analyses of efficient survey strategies for the detection of fast transients \citep[e.g.,][]{Bianco_2019,Andreoni_2019,SeBi2019}. We aim to perform a realistic and comprehensive analysis of all of the different factors that go into the generation of an observing plan, so as to construct concrete recommendations for such kilonova searches. We include factors that cannot easily be taken into account in analytic explorations of the same nature, and as a result provide more realistic predictions on rate constraints and the relative improvement between different strategies.

We describe a path-finder approach to improving searches for serendipitous multi-messenger sources with wide FOV survey systems, focusing on ZTF. The efficiency of the search for fast transients depends on many factors, including cadence, filter choice, sky coverage, and depth of observations \citep{AnCo2019}. We demonstrate that our strategy for survey cadence makes it possible to detect these sources more efficiently than has been possible until now.
Given that the rates and types of background transients are generally known, using ZTF as an example, we can make predictions of the number of transients that will require follow-up to perform characterization and classification.

Our paper is structured as follows: we start by describing the scheduling of these observations and their coverage of multi-messenger events in Section~\ref{sec:observing}; we then discuss the efficiency of counterpart detection for each of these strategies in Section~\ref{sec:results}, and summarize our conclusions and future outlook in Section~\ref{sec:summary}.

\section{Simulated Observing Plans}
\label{sec:observing}

We aim to determine a survey strategy for ZTF that maximizes the probability of detecting a serendipitous kilonova. 
During ZTF's Phase I, which lasted from March 2018 to October 2020, the telescope time was split up between several different programs \citep{Bellm:19:ZTFScheduler}.
Forty percent of the telescope time has been used for public surveys supported by an NSF Mid-Scale Innovations Program (MSIP); another forty percent has been used by the ZTF partnership; and the final twenty percent was disbursed by the Caltech Time Allocation Committee. In ZTF Phase II, fifty percent of the telescope time will be used for public surveys---primarily a two-day cadence survey of the Northern Hemisphere Sky---thirty percent for the ZTF partnership, and twenty percent for the Caltech TAC.

\begin{figure}
\begin{center}
 \includegraphics[width=\columnwidth]{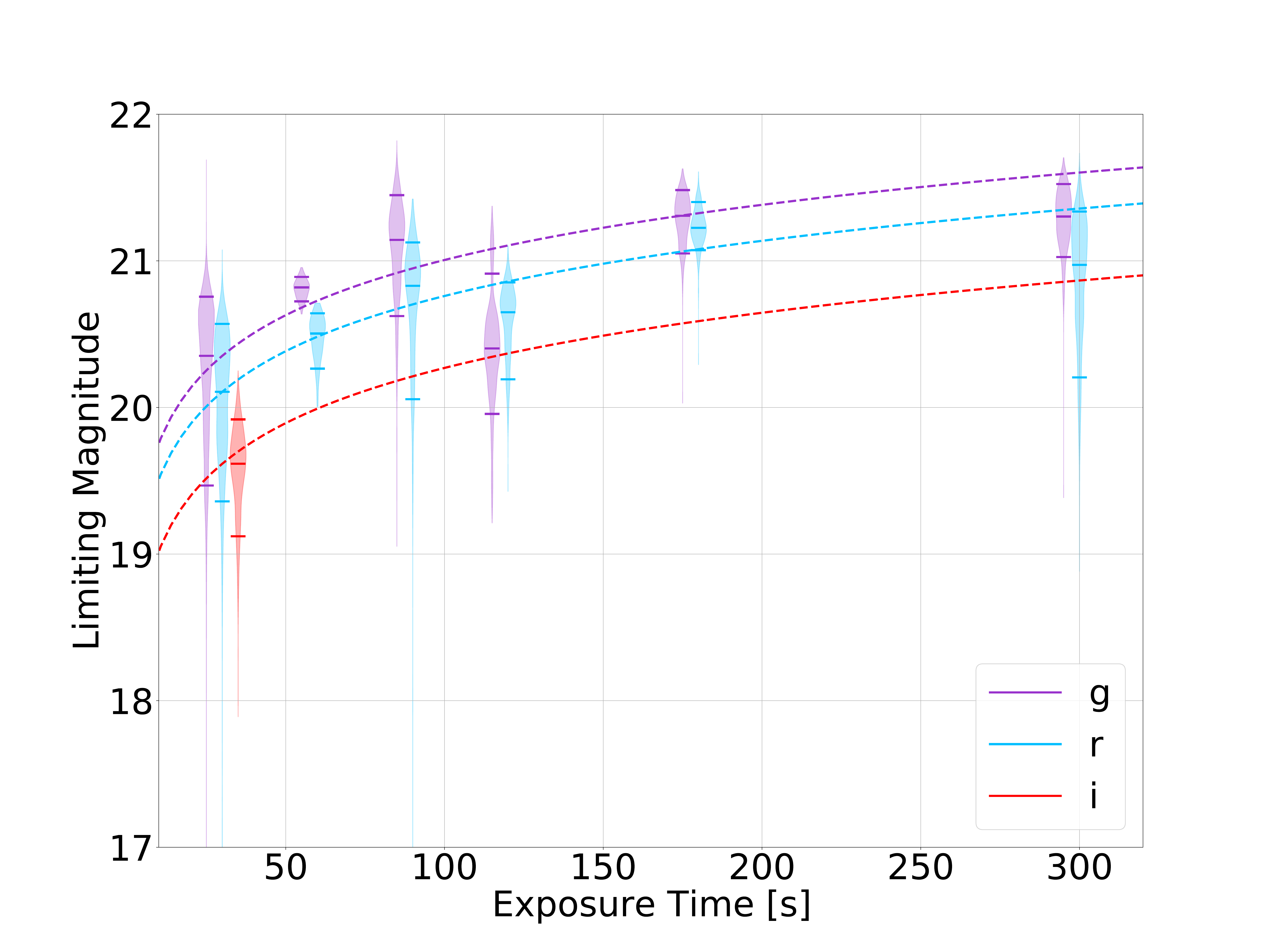}
  \caption{Probability density, shown in the form of ``violins'', for the limiting magnitudes in $g$-, $r$-, and $i$-bands (represented by purple, light blue, and red colors respectively) for the exposure times employed during ZTF Phase I. The three markers represent the 10th, 50th, and 90th percentiles respectively, and the extended tails are caused predominantly by bad weather and high airmass observations. No conditions have been imposed on factors such as airmass, moon phase, and seeing. Also shown in dashed lines are the expected limiting magnitudes as a function of the median 30\,s exposures in each passband using the expected $\sqrt T$ scaling (where $T$ is exposure time) appropriate for observations dominated by the sky background. Note that the $g$-, $r$-, and $i$-band violins are slightly shifted from each other for ease of interpretation.}
 \label{fig:limmag}
\end{center}
\end{figure}


We adopt a simplified scheme in our comparison to ZTF's Phase I cadence. In ZTF's Phase II, 50\% of the night will be dedicated to an all-sky survey, covering the visible sky every two nights, in 30\,s exposures. We denote this survey as the ``Nominal Survey''. We assume that the remaining 50\% of the night is available for survey optimization. Although this is not completely feasible for a single program, it will nevertheless be informative as scheduling strategies can be adopted to target particular science cases. We denote this survey as the ``Kilonova Survey'', during which we also avoid any observations in the galactic plane. In our analysis, we will systematically change the exposure time, the filters, and the cadence in order to optimize observing strategies targeting kilonovae. In essence, the half of the night falling under the Nominal Survey follows the 30\,s exposure program, with $g$- and $r$-band epochs for each field, while the half that falls under the Kilonova Survey is varied accordingly.

\begin{figure*}[t]
 \includegraphics[width=\linewidth]{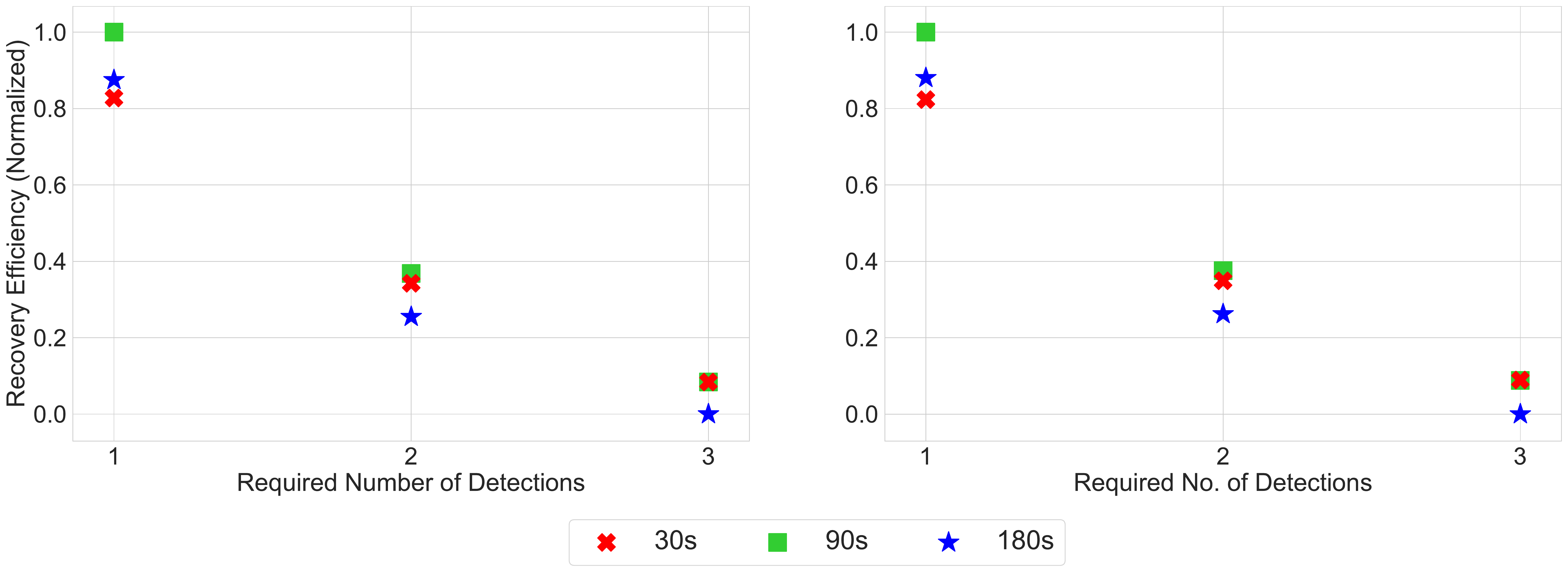}
 \caption{Recovery efficiencies plotted for different detection requirements, assuming regular weather conditions. On the left, we show results for the kilonova population with viewing angles uniformly distributed in cosine, and on the right, for the population with all kilonovae having the same viewing angle as AT2017gfo. The 30\,s observing plan is represented by red crosses, 90\,s by green squares, and 180\,s by blue stars.}
  \label{fig:effs_sine_vs_fixed}
\end{figure*}


In order to generate realistic schedules to serve as representatives for each strategy, we use \texttt{gwemopt} \citep{CoTo2018,CoAn2019,AlCo2020}, a code-base that was originally designed to perform Target-of-Opportunity scheduling, but modified here to suit our purposes. We generate realistic schedules, taking into account factors such as telescope configuration and observational/diurnal constraints. Additionally, in order to generate realistic simulations, the limiting magnitudes in the generated schedules have been computed based on past ZTF observations. We show in Figure~\ref{fig:limmag} the limiting magnitudes for $g$-, $r$-, and $i$-bands for the exposure times employed during ZTF Phase I. We also show in dashed lines the expected limiting magnitudes as a function of the median 30\,s exposures in each passband, using the expected $\sqrt T$ scaling (where $T$ is exposure time) appropriate for observations dominated by the sky background, showing the observations' consistency with the expected evolution. There is some deviation from the dashed lines, likely due to the fact that we did not filter for any specific range of factors (e.g., airmass, moon phase, weather conditions), also meaning that we are able to sample a wide variety of conditions that present a more realistic depiction of ZTF observations.

Based on the different elements that constitute the structure of an observing plan (e.g., exposure time and filters), we can use different models to assess a given strategy's performance. 
In the following, we will use \texttt{simsurvey} \citep{FeNo2019}, a software package that simulates the expected lightcurves for the transient. Based on such different factors and setups, we can then determine the possibility of a transient discovery.

The simulated lightcurves used to calculate efficiencies in Section~\ref{sec:results} are extracted from a kilonova model with two ejecta mass components \citep{Bul2019,Dietrich:2020lps}; we choose values of $0.005M_\odot\,$ and $0.05M_\odot$ for the dynamical ($M_{\rm ej,dyn}$) and post-merger wind ($M_{\rm ej,pm}$) ejecta masses respectively, and $45^{\circ}$ for the half-opening angle of the lanthanide-rich component.
These intrinsic parameters were chosen as a best fit for the AT2017gfo lightcurve \citep{Dietrich:2020lps}. To explore results for different viewing angle assumptions, we simulate two different kilonova populations: one with viewing angles uniformly distributed in cosine, and one with all kilonovae possessing a fixed viewing angle chosen to be $30^{\circ}$, i.e., similar to AT2017gfo.

\section{Results}
\label{sec:results}

\subsection{Exposure Time}
\label{subsec:exposure_times}
We first aim to produce an exposure-time optimized strategy, keeping the total time allowed for observations constant so as to fairly evaluate the performance of each plan. We will explore how longer exposure times affect kilonova rate constraints, which can in turn constrain rates of binary neutron star and neutron-star black-hole mergers. There is a natural optimization where the longer exposure times achieve more depth, but reduce the number of observations that can be scheduled for the Kilonova Survey, so the interplay between depth achieved and area covered dictates the outcome of this analysis. Here, we focus on 30\,s, 90\,s, and 180\,s exposures. We limit to 180\,s exposures, since -- as we will soon show -- longer exposure times do not prove suitable for surveys that aim to cover such large sky areas.

We can first set some rough expectations for the results by approximating the relative ``volumes'' that each exposure time is sensitive to, assuming all-sky coverage and a fixed total observing time. From Figure~\ref{fig:limmag}, we expect that 30\,s exposures will yield a limiting magnitude of $r$\,$\sim$\,20.1\,mag. Since we are assuming that the sensitivity scales as $\sqrt T$, the limiting magnitudes for the 90\,s and 180\,s exposures are $\sim$\,20.7 and 21.1\,mag respectively. From there, we can compute the approximate sensitive volumes for each of the exposures assuming a transient of $M_{r}$ = $-$16\,mag (which is approximately the peak absolute magnitude of AT2017gfo in the $r$-band). Incorporating the ZTF declination limit of $\sim$\,$-30^{\circ}$, we obtain volumes of 0.014, 0.033, and 0.057 Gpc$^3$ for 30, 90, and 180\,s exposures respectively. In addition, taking into account ZTF's overhead per exposure of $\sim$\,10\,s, we calculate the \textit{total} volume observable in 30\,s exposures within the time it takes for a 90\,s and 180\,s exposure to complete; these volumes are $\sim$0.035 Gpc$^3$, and $\sim$0.067 Gpc$^3$ respectively. While the 90\,s exposure and equivalent 30\,s exposures cover similar volumes, the 30\,s exposures have a clear advantage over the 180\,s exposure in their volume coverage.
These calculations can also be done for surveys other than ZTF. We must note an additional caveat in this simplified analysis: these numbers assume distinct fields, i.e., no consecutive visits of the same area of the sky, whereas the schedules used in this section contain $g$- and $r$-band epochs for each field. Nevertheless, we can now compare these rough expectations to the actual results from \texttt{gwemopt} and \texttt{simsurvey}, discussed below.



We generate three schedules that adopt 30\,s, 90\,s, and 180\,s exposures respectively for the Kilonova Survey observations, with both $g$- and $r$-band epochs scheduled per night, with a cadence of $\sim$\, one to two nights. The schedules span one year, covering from 2019-01-01 to 2020-01-01, in order to understand the expected detection prospects over realistic program lengths.





The resulting normalized recovery efficiencies are plotted in Figure~\ref{fig:effs_sine_vs_fixed} for different detection requirements, both for the viewing angle population that is uniformly distributed in cosine (left plot) and the population with a fixed viewing angle of $30^{\circ}$ (right plot). In practice, fade (or rise) rates, found by performing linear fits before/after the brightest detection, are an essential determinant of whether a detected transient is a possible kilonova \citep{andreoni2020constraining}. Therefore, while all cases require further follow-up to confirm kilonovae, the single-epoch case in particular does not provide enough useful information about the nature of the transient. Requiring two to three detections of at least 5-$\sigma$ significance is thus standard in the filtering process \citep{SeBi2019,andreoni2020constraining}, making it more likely to identify the expected rapid evolution of kilonovae. For the results in Figure~\ref{fig:effs_sine_vs_fixed}, we computed the observations' limiting magnitudes taking into account all possible weather scenarios for ZTF, including nights when the dome is closed due to bad weather; we denote these schedules as those representing ``regular seeing conditions''. 

\begin{table}
\centering
\caption{Constraints on the kilonova rate (assuming that no kilonovae were detected throughout the execution of each schedule) when imposing a two-detection requirement, for different seeing conditions, exposure times, and simulated kilonova populations.}
\label{tab:exptime_rates}
\begin{tabular}{lccr} 

\begin{scriptsize}\textbf{Viewing Angle Distribution}\end{scriptsize} & \begin{scriptsize}\textbf{Exposure Time}\end{scriptsize} & \begin{scriptsize} \textbf{Kilonova Rate Upper Limit} \end{scriptsize} \\
& \begin{scriptsize} (seconds) \end{scriptsize} & \begin{scriptsize} (Gpc$^{-3}$yr$^{-1}$) \end{scriptsize}\\
\hline
\multicolumn{3}{c}{\textit{Regular Seeing Conditions}} \\
\hline
\begin{scriptsize} \textbf{Cosine}   \end{scriptsize}& \begin{scriptsize} $30$ \end{scriptsize}  &\begin{scriptsize} $3529$ \end{scriptsize} \\
\begin{scriptsize}   \end{scriptsize}& \begin{scriptsize} $90$ \end{scriptsize}  &\begin{scriptsize} $3409$ \end{scriptsize} \\
\begin{scriptsize}   \end{scriptsize}& \begin{scriptsize} $180$ \end{scriptsize}  &\begin{scriptsize} $4054$ \end{scriptsize} \\
\hline
\begin{scriptsize} \textbf{Fixed at} $\mathbf{30^{\circ}}$  \end{scriptsize}& \begin{scriptsize} $30$ \end{scriptsize}  &\begin{scriptsize} $1807$ \end{scriptsize} \\
\begin{scriptsize}    \end{scriptsize}& \begin{scriptsize} $90$ \end{scriptsize}  &\begin{scriptsize} $1734$ \end{scriptsize} \\
\begin{scriptsize}    \end{scriptsize}& \begin{scriptsize} $180$ \end{scriptsize}  &\begin{scriptsize} $2068$ \end{scriptsize} \\
\hline
\multicolumn{3}{c}{\textit{Good Seeing Conditions}}  \\
\hline
\begin{scriptsize} \textbf{Cosine}   \end{scriptsize}& \begin{scriptsize} $30$ \end{scriptsize}  &\begin{scriptsize} $2113$ \end{scriptsize} \\
\begin{scriptsize}   \end{scriptsize}& \begin{scriptsize} $90$ \end{scriptsize}  &\begin{scriptsize} $2069$ \end{scriptsize} \\
\begin{scriptsize}   \end{scriptsize}& \begin{scriptsize} $180$ \end{scriptsize}  &\begin{scriptsize} $2362$ \end{scriptsize} \\
\hline
\begin{scriptsize} \textbf{Fixed at} $\mathbf{30^{\circ}}$  \end{scriptsize}& \begin{scriptsize} $30$ \end{scriptsize}  &\begin{scriptsize} $1095$ \end{scriptsize} \\
\begin{scriptsize}    \end{scriptsize}& \begin{scriptsize} $90$ \end{scriptsize}  &\begin{scriptsize} $1064$ \end{scriptsize} \\
\begin{scriptsize}    \end{scriptsize}& \begin{scriptsize} $180$ \end{scriptsize}  &\begin{scriptsize} $1215$ \end{scriptsize} \\
\end{tabular}
\end{table}
 
The 90\,s exposures perform very well for the one-detection requirement, but for the two- and three-detection requirements, their dominance over 30\,s exposures significantly lessens. In the left sub-plot of Figure~\ref{fig:effs_sine_vs_fixed}, we see improvements of 15\%, 4\%, and 0.06\% for the one-, two-, and three-detection requirements respectively; in the right plot, these correspond to improvements of 15\%, 4\%, an $-$0.3\% respectively. The 180\,s exposures clearly do not fare as well for any of the realistic filtering requirements. Assuming ``good'' seeing conditions (i.e., only sampling from past ZTF observations that reach within the top 50\% of limiting magnitudes), for the two-detection requirement, the improvement from 30\,s to 90\,s exposures corresponds to 2.5\% and 3\% for the viewing angle population that is uniformly distributed in cosine, and the population with a fixed viewing angle of 30$^{\circ}$, respectively. Therefore, overall, the 90\,s exposures mostly help in detecting a larger portion of the simulated kilonova population (represented by the one-detection requirement), but are only slightly more beneficial than 30\,s exposures when requiring multiple detections in the lightcurve of any given kilonova.

\begin{figure}
\begin{center}
 \includegraphics[width=\columnwidth]{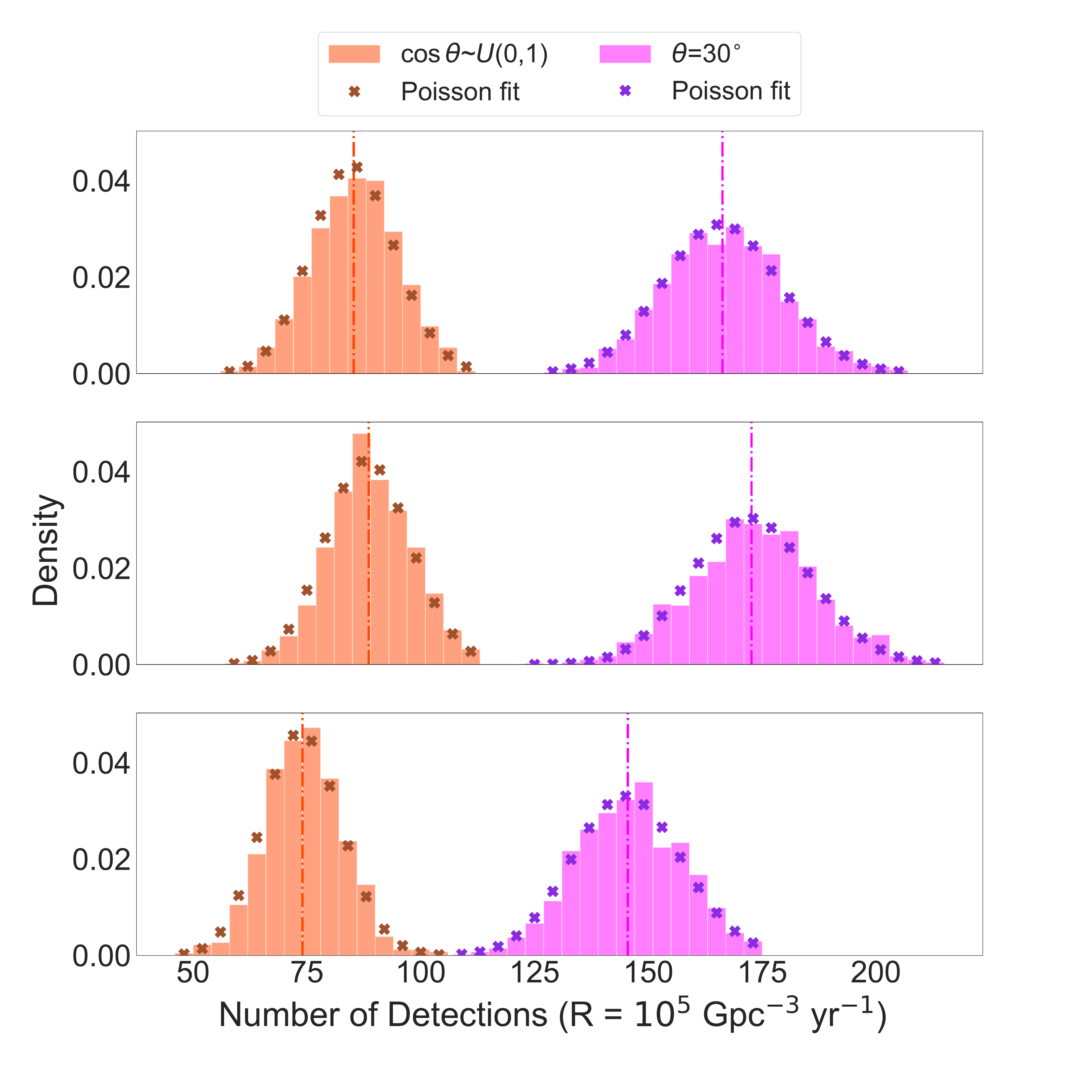}
  \caption{Distribution of detections (at a rate $R$ = $10^{5}$ Gpc$^{-3}$ yr $^{-1}$) for schedules employing 30\,s (top subplot), 90\,s (middle subplot), and 180\,s (bottom subplot) exposure times respectively; these schedules assume regular conditions, and we impose a two-detection requirement as part of the filtering criteria. The orange histograms show results for the kilonova population with viewing angles uniformly distributed in cosine, and the magenta histograms show results for the kilonova population with $\theta$ fixed at  $30^{\circ}$; the mean of each distribution is plotted as a vertical dashed line in the same color. The Poisson fits for each of the datasets are plotted as brown and purple crosses respectively.}
 \label{fig:rate_dists}
\end{center}
\end{figure}

In the case that no kilonova is found throughout the execution of the year-long observing plan, as in past observing campaigns, we can constrain the kilonova rate accordingly. We do this by first simulating the kilonova populations 1000 times for each observing plan, so as to obtain a large sample of the number of detections at an arbitrary rate; as shown in Figure~\ref{fig:rate_dists}, these samples follow a Poisson distribution. From the mean, we are then able to extract the linear relationship between the rate and number of detections. As done in \cite{andreoni2020constraining}, for a 95\% confidence interval, the upper limit on the rate is defined as the rate at which three simulated kilonovae pass the filtering criteria for the respective observing plan. The derived rates are shown in Table~\ref{tab:exptime_rates}. For regular seeing conditions, and with viewing angles following a distribution that is uniform in cosine, the kilonova rate constraint improves from $R < 3529$ Gpc$^{-3}$ yr$^{-1}$ to $R < 3409$ Gpc$^{-3}$ yr$^{-1}$ when using 90\,s exposures instead of the nominal 30\,s exposures for the Kilonova Survey.


Going back to the relative volumes that were derived, we can see that for the one- and two-detection requirements, our conclusions slightly deviate from our earlier predictions since 90\,s exposures perform better than the 30\,s exposures. We can reason this by noting that the relative volume constraints harbor an initial assumption of M = $-16$\,mag for the transient. This is only the peak magnitude of the $r$-band lightcurve (in the $30^{\circ}$ viewing angle case), so observations before or after that point would benefit from deeper exposures; in addition, the $g$-band lightcurve peaks at around 0.3\,mag fainter than the $r$-band lightcurve (for a $30^{\circ}$ viewing angle), and fades rapidly, reaching M = $-$12.3\,mag just one day after peak. Therefore, slightly deeper observations than the nominal 30\,s exposures can be helpful, especially in the 90\,s case, since there is not much loss in coverage either.

It is important to note that this result can contrast with optimal strategies for Target of Opportunity observations of GW and GRB events, for which the areas of interest are constrained to much more reasonable sky localizations; in such cases, the gain in depth from longer exposure times ($\gtrsim$\,180\,s) can usually be exploited without having to worry too much about the loss in coverage \citep[e.g.,][]{GhCh2017,CoDi2020}.


\subsection{Filters}
\label{subsec:filters}
Due to the fact that kilonovae rapidly redden with time \citep{Ta2016, KaKy2016}, the inclusion of near-infrared filters tends to increase the chances of kilonova detection; $i$-band observations have been suggested to be of great benefit for ZTF's kilonova searches \citep{Andreoni_2019,AnCo2020,carracedo2020detectability}.

We therefore generated another year-long schedule, this time implementing $g$-, $r$- and $i$-band epochs per field in a night (all of which adopt 90\,s exposures), with an approximately 1-2 day cadence, for the Kilonova Survey observations. By comparing the total number of kilonovae recovered (out of the 10,000 injected) based on their detections in each band, we can then infer the possible benefits of redder $i$-band observations.

\begin{figure}
 \includegraphics[width=\columnwidth]{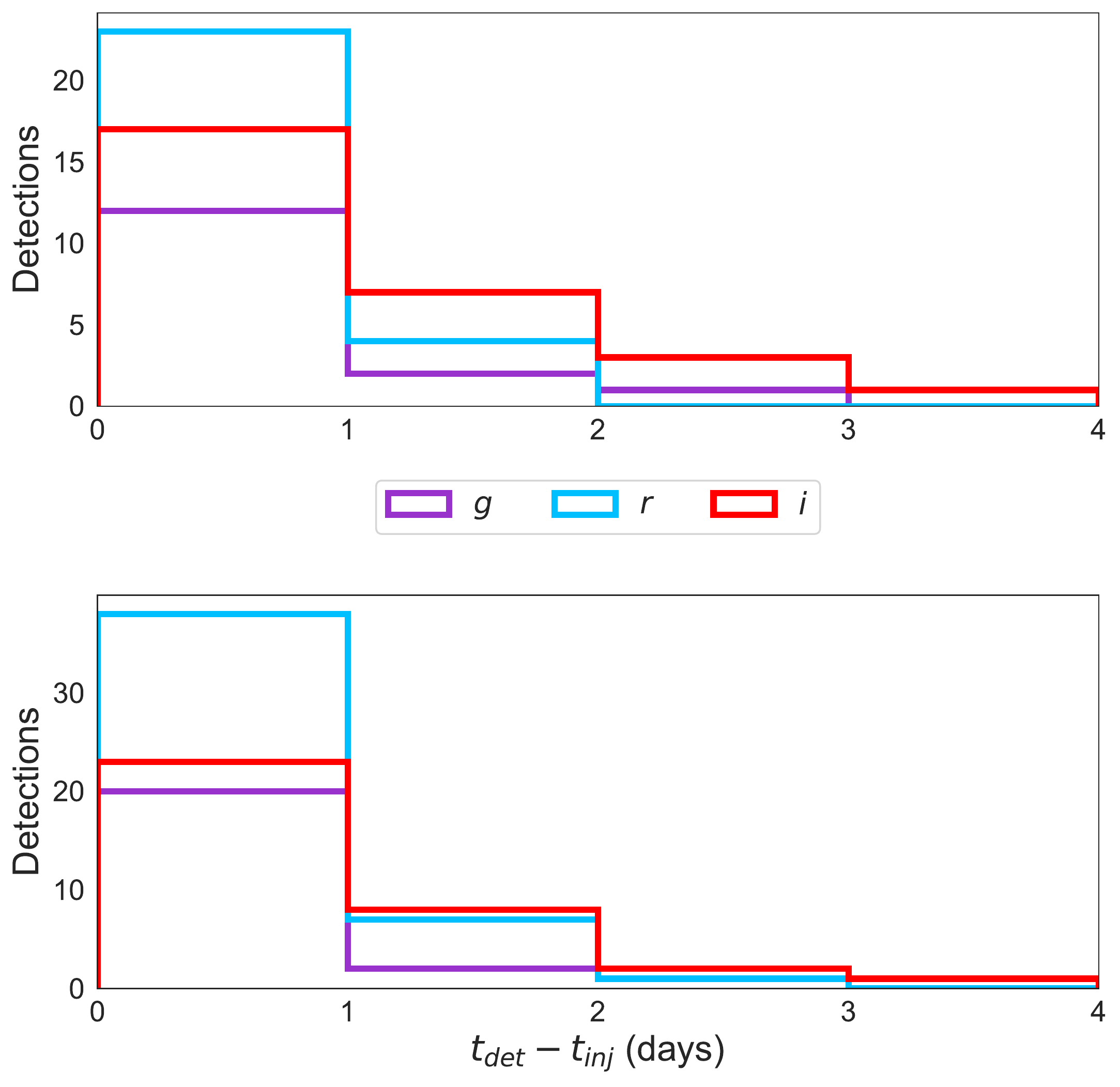}
  \caption{Number of detections, out of 10,000 injections, in $g$, $r$, and $i$ bands with respect to the phase of the kilonova at time of detection, assuming regular weather conditions. The purple, light blue, and red colors represent the $g$, $r$, and $i$ bands respectively. Here we show results adopting both the distribution of viewing angles uniform in cosine (top) and the AT2017gfo-like viewing angles (bottom). Detections in the $i$-band start to dominate those in $g$- and $r$- bands $\gtrsim$\,1 day post-merger in both cases.}
 \label{fig:iband_pdets}
\end{figure}

Since kilonovae fade more slowly in redder bands, the benefit of adopting such a strategy can first be visualized through Figure~\ref{fig:iband_pdets}, in which all of the detections in each of the filters are plotted as a function of phase. The $i$-band detections extend much further in terms of the time at which the kilonova was detected relative to the merger time, and so are vital to identifying kilonovae at later times.

\begin{figure}
 \includegraphics[width=\columnwidth]{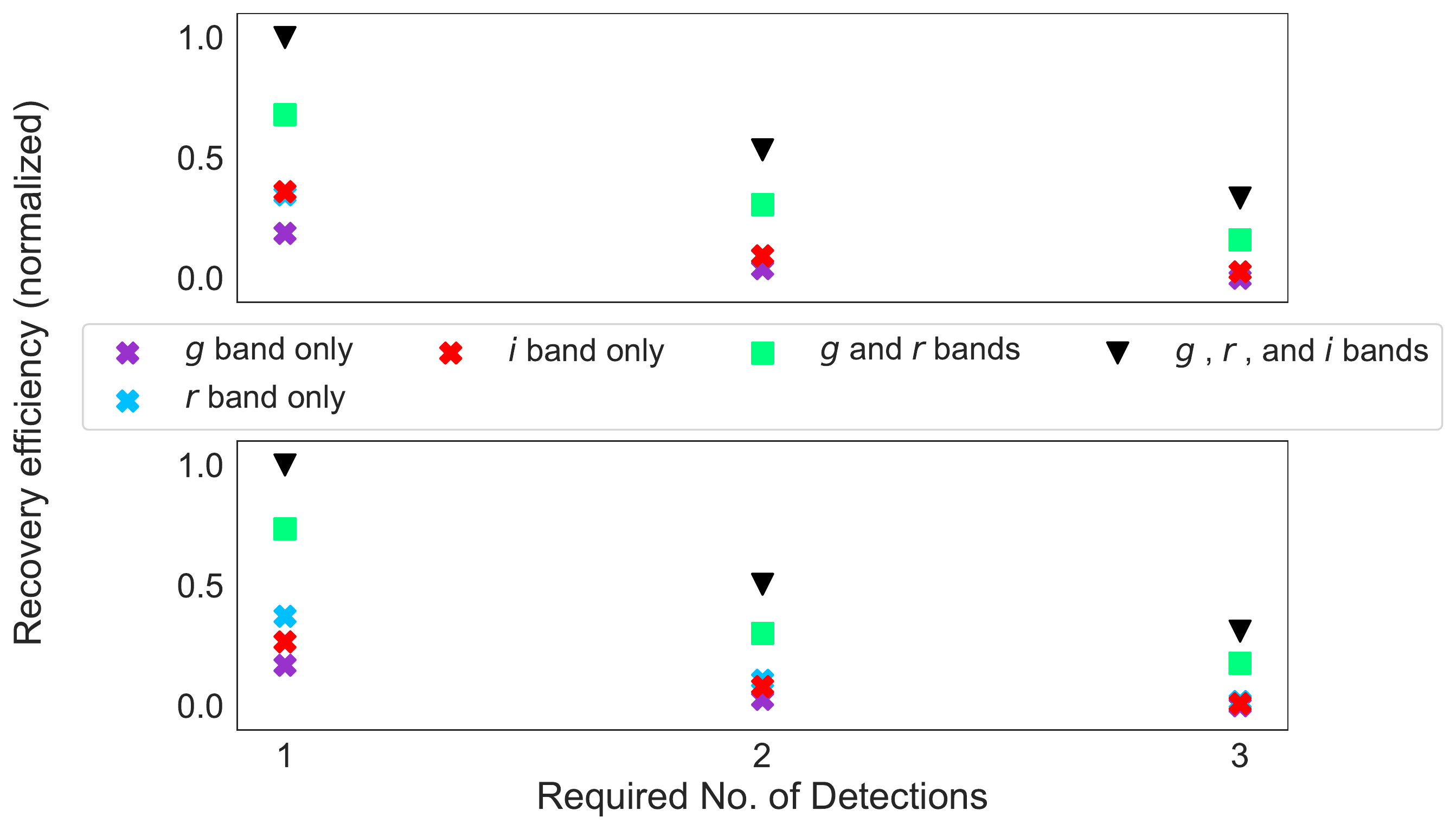}
  \caption{A breakdown by filter of the recovered kilonovae for a year-long schedule adopting 90\,s $g$-, $r$-, and $i$- band exposures (for the Kilonova Survey); we show results assuming regular conditions, adopting both the distribution of viewing angles uniform in cosine (top) and the AT2017gfo-like viewing angles (bottom). The focus here is the improvement when including the $i$-band detections (shown as black triangles) as opposed to only those in the $g$- and $r$-bands (shown as bright green squares). The $g$, $r$, and $i$ bands are represented by purple, light blue, and red crosses respectively.}
 \label{fig:iband_detectedKNe}
\end{figure}

More holistically, we can investigate the efficiency of kilonova recovery in different bands for a number of detection requirements, as shown in Figure~\ref{fig:iband_detectedKNe} for regular viewing conditions. We are clearly able to recover a much higher number of kilonovae by including $i$-band observations (indicated in black) in addition to those just in $g$- and $r$-bands (indicated in purple). The top subplot shows results for the kilonova population with viewing angles uniformly distributed in cosine, and the bottom subplot for the AT2017gfo-like viewing angle population. We find that $i$-band observations are especially helpful for the kilonova population with viewing angles uniformly distributed in cosine, and become more beneficial as the detection requirements become more rigorous. More concretely, we see improvements of $\sim$ 46\%, 71\%, and 100\% for the one-, two-, and three-detection requirements respectively, from the recoveries in just $g$- and $r$- bands. Results for the fixed viewing angle population are broadly consistent, leading to increases of $\sim$ 34\%, 61\%, and 63\% for each of the one-, two-, and three-detection requirements. Similar to the results in Section~\ref{subsec:exposure_times}, the different detection requirements yield different numbers of kilonovae. For example, in the $g$-, $r$-, and $i$-band simulation sets, moving from one detection to two detections loses $\sim$25\% of the recoveries, and one to three detections loses $\sim$50\%.

We can also derive rate constraints using these $g$-,$r$, and $i$-band observing plans we have generated, to further quantify the benefit of scheduling redder bands. As done in Section~\ref{subsec:exposure_times}, we will assume that there were no successful serendipitous kilonova discoveries throughout the whole year of observing, and derive constraints on the kilonova rate based on a two detection requirement. For the most realistic scenario (regular seeing conditions, and a uniform-in-cosine distribution of viewing angles for the simulated kilonova population) we obtain $R < 2749$ Gpc$^{-3}$yr$^{-1}$. This rate is to be compared to $R < 3529$ Gpc$^{-3}$yr$^{-1}$ for the 30\,s $g$- and $r$-band schedule in Section~\ref{subsec:cadence}, and $R < 3409$ Gpc$^{-3}$yr$^{-1}$ for the 90\,s $g$- and $r$-band schedule. Rate constraints for other scenarios are shown in Table~\ref{tab:iband_rates}. Therefore, such a strategy is beneficial even when no kilonovae are detected whatsoever, in that we are able to obtain significantly tighter constraints on the kilonova rate. It is to note that we also have not completely controlled how often these observations are conducted for a given field night-to-night (to more than within a few days), which would likely yield even tighter constraints on the kilonova rate.

\begin{table}
\centering
\caption{Constraints on the kilonova rate for the schedules employing $i$-band observations, in addition to the usual $g$- and $r$-band exposures, assuming that no kilonovae were detected throughout the execution of each schedule. We use the method in Section~\ref{subsec:exposure_times} to compute these upper limits, and impose a two-detection requirement.}
\label{tab:iband_rates}
\begin{tabular}{lccr}
\begin{scriptsize}\textbf{Viewing Angle Distribution}\end{scriptsize} & \begin{scriptsize} \textbf{Kilonova Rate Upper Limit} \end{scriptsize} \\
& \begin{scriptsize} (Gpc$^{-3}$yr$^{-1}$) \end{scriptsize}\\
\hline
\multicolumn{2}{c}{\textit{Regular Seeing Conditions}} \\
\hline
\begin{scriptsize} \textbf{Cosine}   \end{scriptsize} &\begin{scriptsize} $2749$ \end{scriptsize} \\
\begin{scriptsize} \textbf{Fixed at} $\mathbf{30^{\circ}}$  \end{scriptsize} &\begin{scriptsize} $1508$ \end{scriptsize} \\
\hline
\multicolumn{2}{c}{\textit{Good Seeing Conditions}}  \\
\hline
\begin{scriptsize} \textbf{Cosine}   \end{scriptsize} &\begin{scriptsize} $1617$ \end{scriptsize} \\
\begin{scriptsize} \textbf{Fixed at} $\mathbf{30^{\circ}}$  \end{scriptsize} &\begin{scriptsize} $893$ \end{scriptsize} \\
\end{tabular}
\end{table}

\subsection{Cadence}
\label{subsec:cadence}
Aside from exposure times and filters, cadence is another essential determinant of the optimal survey strategy. Naturally, due to the fast-evolving nature of kilonovae, high-cadence strategies are important to optimize their detection with the necessary color and brightness information (see \cite{Andreoni_2019} for a more in-depth discussion of cadence-optimized strategies for detecting kilonovae). In order to probe this aspect in a practical manner, we generate a two-week-long 300\,s schedule ($g$- and $r$-band exposures), injecting 10,000 kilonovae into \textit{each field}, and computing the per-field efficiencies based on the schedule. We choose 300\,s exposures for this part of the analysis because longer exposure times yield much higher percent recoveries in comparison to shorter exposure times when they are computed on a per-field basis (i.e., total sky coverage is no longer a factor); this increase leads to less fluctuation in the results, and so helps in isolating cadence from other confounding variables. 

\begin{figure}
 \includegraphics[width=\columnwidth]{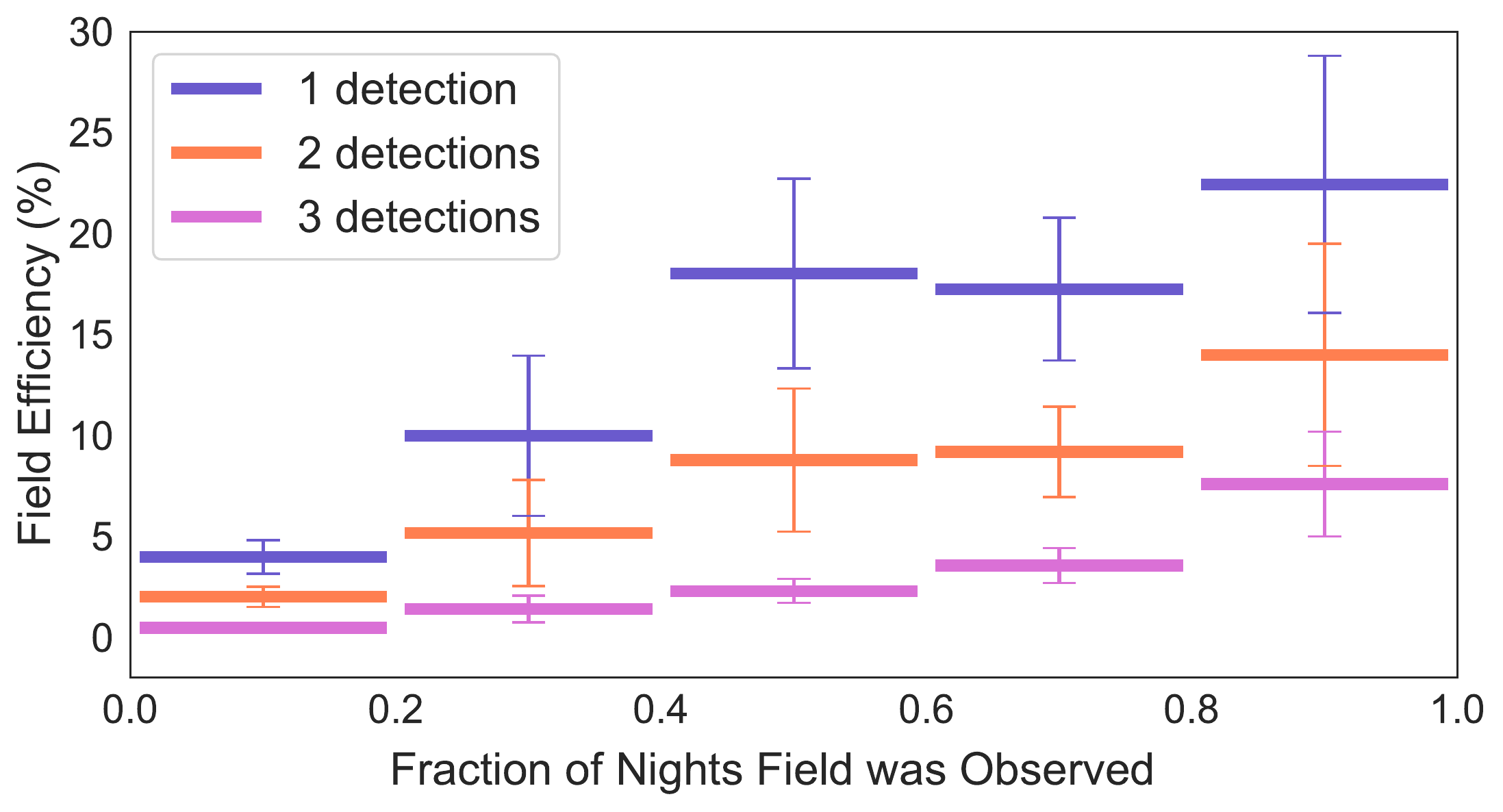}
  \caption{Field efficiencies binned by the fraction of nights the field was observed (e.g., a field that was observed every night would correspond to a value of 1.0, or a nightly cadence, and a field observed every other night would correspond to a fraction of 0.5, or a 2-night cadence) for one-, two-, and three-detection requirements (shown in dark purple, orange, and pink colors respectively). We only take into account the $300$\,s  Kilonova Survey observations. For all filtering criteria, there is a clear preference for fields observed at a higher cadence.}
 \label{fig:cadence_numnights}
\end{figure}

In Figure~\ref{fig:cadence_numnights}, we show the field efficiencies as a function of the fraction of nights during which they were observed, taking into account the Kilonova Survey observations only. We use the AT2017gfo-like viewing angle population. There is some fluctuation present in the results, mostly evident for the one-detection requirement, as shown through the error bars representing one sigma standard deviation. In addition, multiple observations within close proximity to each other (on the order of $\sim$\,days) are a less important factor when employing such a lenient filtering requirement.

We can nevertheless see that, for all filtering requirements, there is a positive correlation between the two variables. For example, the median number of detected kilonovae for fields observed 80-100\% of nights (i.e., with a close to nightly cadence), as compared to that for fields observed 20-40\% of nights (a cadence of $\sim$ 3-5 days), increases by around 171\% and $\sim$ 470\% respectively for the two- and three detection requirements. 
\begin{figure}
 \includegraphics[width=\columnwidth]{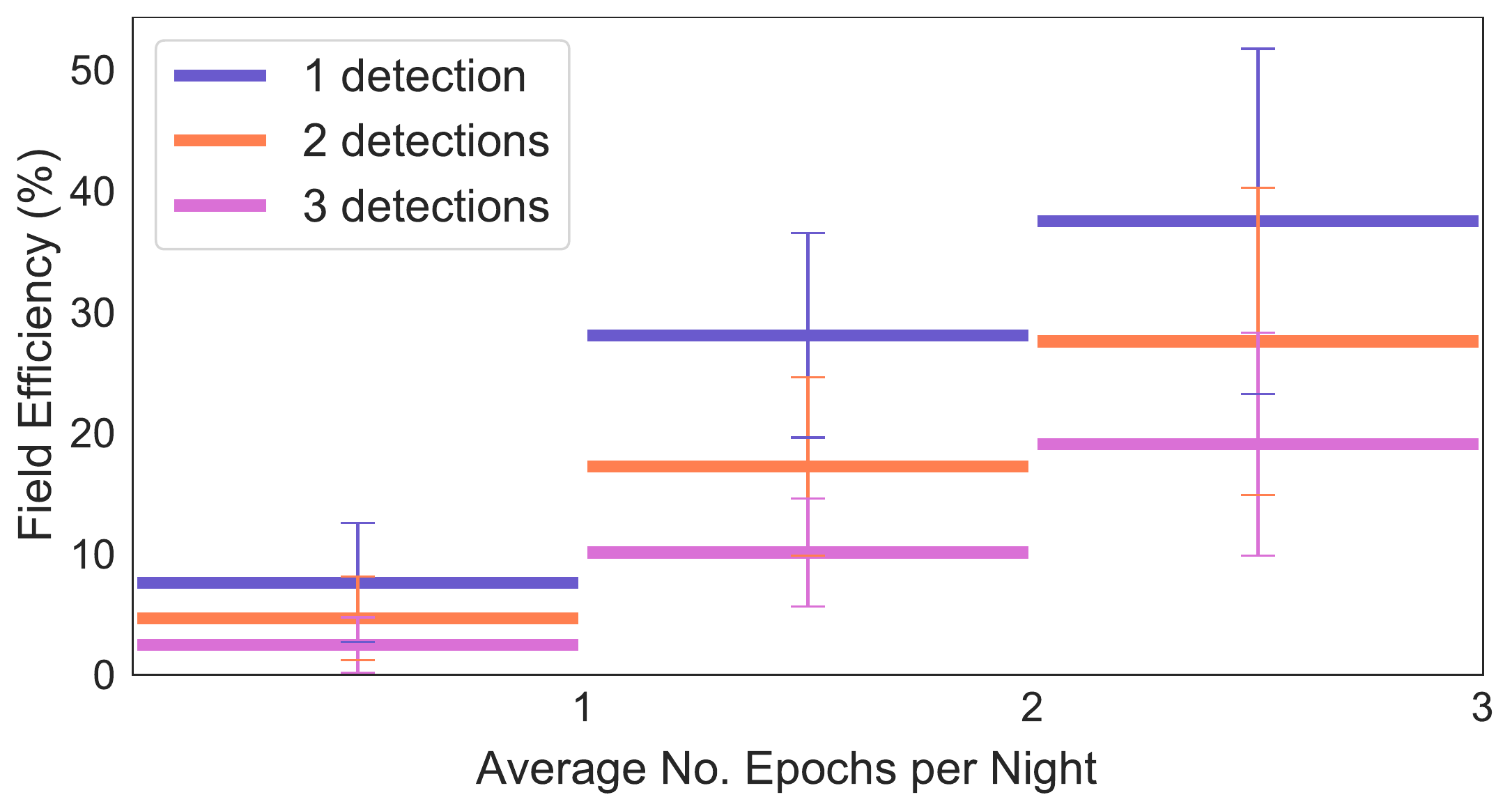}
 \caption{Field efficiencies binned by the average number of epochs per night, allowing up to three epochs per night ($g$-, $r$- and $i$-band) to be scheduled. Results are shown for one-, two-, and three-detection requirements, represented by dark purple, orange, and pink colors respectively.}
 \label{fig:cadence_numepochs}
\end{figure}

\begin{figure*}
 \includegraphics[width=\textwidth]{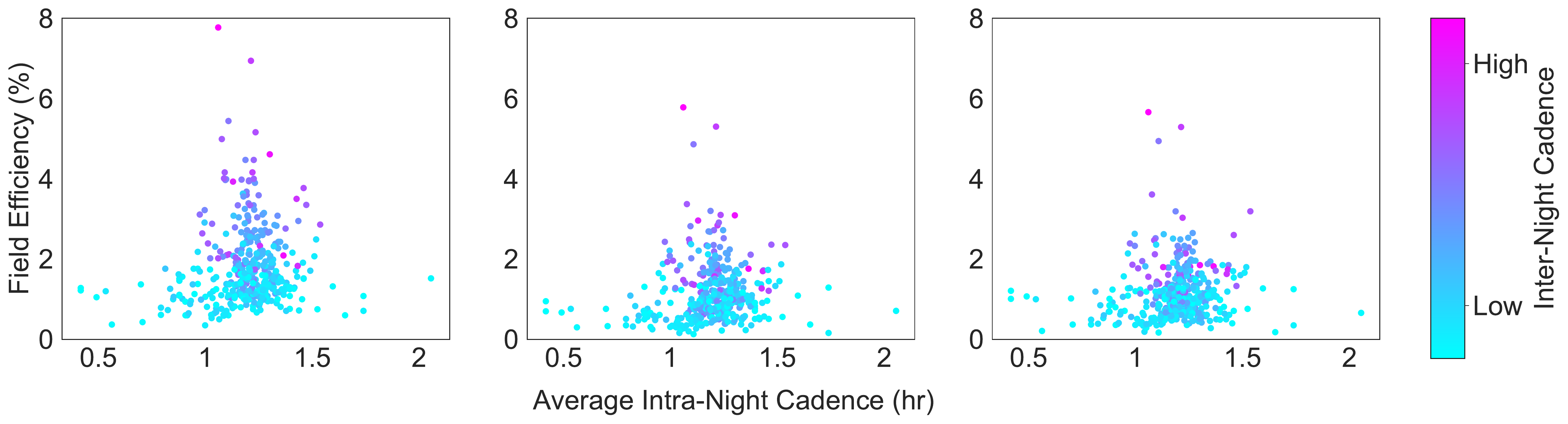}
 \caption{Field efficiency as a function of average intra-night cadence (i.e., time between observations in different filters), with inter-night cadence (how often the field is re-visited night to night) as a color bar, for different filter combinations: $g$-$r$-$i$ on the left, $g$-$i$-$g$ in the middle, and $g$-$i$-$i$ on the right. We can see that the $g$-$r$-$i$ combination outperforms the others in recovering the population of simulated kilonovae, especially for fields with a close to nightly inter-night cadence.}
 \label{fig:mega_cadence}
\end{figure*}

Running another 2-week schedule in which we allowed up to three epochs to be scheduled per night ($g$-, $r$-, and $i$-band), the field efficiencies were then binned according to the average number of exposures per night in Figure~\ref{fig:cadence_numepochs}. Due to the fact that $i$-band exposures are included, overall field efficiencies here are notably higher than those shown in Figure~\ref{fig:cadence_numnights} (in which case only $g$- and $r$- bands were scheduled). We see increases of 60\% and 88\% in the number of kilonovae detected when an average of 2-3 epochs are scheduled, rather than 1-2 epochs, for the two- and three-detection requirements respectively.

We can also explore the benefit of different filter combinations in the search for fast-evolving transients, as has been done in \citet{Andreoni_2019} and \citet{Bianco_2019} for the Vera Rubin Observatory. Here, we aim to probe this in a more practical manner, utilizing the year-long $g$-, $r$-, and $i$-band schedule from Section~\ref{subsec:filters}, and using the exact same observing plan but with different filter combinations to explore their effect on the detectability of kilonovae. To visualise the distribution of efficiencies across all observations for each combination, we have plotted field efficiencies as a function of average time between observations in the same night, with inter-night cadence shown as a colorbar, in Figure~\ref{fig:mega_cadence}. We used three different filter combinations: $g$-$r$-$i$ (left sub-plot), $g$-$i$-$g$ (middle subplot), and $g$-$i$-$i$ (right sub-plot), and adopted a two-detection requirement. The $g$-$r$-$i$ filter combination yields the highest field efficiencies, while the $g$-$i$-$g$ and  $g$-$i$-$i$ combinations perform very similarly. Using our scheduling software, most fields are scheduled with an average intra-night cadence between $1$ and $1.5$\,hr, and fields observed at the highest inter-night cadences clearly have the highest efficiencies. If we look at the overall efficiency, the $g$-$r$-$i$ combination has a 55\% higher recovery efficiency than both of the other combinations; for fields within the top 20\% in terms of inter-night cadence, we see a 72\% improvement relative to the $g$-$i$-$g$ combination, and a 61\% improvement relative to the $g$-$i$-$i$ combination. This is likely due to the fact that, for the AT2017gfo-like lightcurves used, there is a much quicker rise in the $r$-band in the first day post-merger. In addition, the peak in the $r$-band is around half a magnitude brighter than in the $g$-band, and the $g$-band lightcurve fades much more quickly. The $g$-$r$-$i$ combination thus increases the likeliness of obtaining at least two detections within the first few days. It is also important to note that ZTF has relatively poorer sensitivity in redder bands, as can be seen in Figure~\ref{fig:limmag}, and this would further decrease the recovery efficiencies in the $g$-$i$-$i$ case than shown; this point is generalizable to other optical facilities as well \citep[e.g.,][]{Oi_2014,ShaMe2015,AbAb2018}. Relying too heavily on $i$-band detections, as is the case in the $g$-$i$-$i$ filter combination, is therefore expected to negatively affect the detectability of the kilonovae; although, of course, the color information given by such combinations can be very useful in identifying kilonovae in practice, assuming that those detections are made \citep{Bianco_2019,Andreoni_2019}.

It is to be expected that higher-cadence strategies are beneficial, but these numbers show that it is prudent even for surveys of this nature -- for which one would expect that covering as much area as possible is the first and foremost goal -- to employ such intra-night multi-epoch strategies.




Since the correlations explored in Figures~\ref{fig:cadence_numnights} and~\ref{fig:cadence_numepochs} are strongest for more rigorous detection requirements, higher-cadence strategies will ensure that more potential kilonovae pass all of the filtering criteria. In addition, adopting a high-cadence survey strategy is very useful in ruling out false positives \citep{Mahabal2018}. Therefore, $\sim$\,three exposures per night, returning to the same field each night, is favorable, and facilitates the detection and identification of such fast transients.

\section{Summary and Conclusions}
\label{sec:summary}

In this study, we have presented an overview of how wide FOV survey strategies such as those used by ZTF may be used to optimize kilonova searches. We assessed the efficiency of detections emerging from these strategies for a number of models, simulating the potential for serendipitous kilonova discovery in realistic conditions. We demonstrated the significant difference in coverage over these timescales with different exposure times and filter combinations.
Finally, we showed how the efficiency of recovery changes as a function of the intra-field cadence.

Having explored the choices that lead to the formulation of the optimal survey strategy in the ZTF search for kilonovae, we may summarize our conclusions as follows:
\begin{description}[font=$\bullet$~\normalfont\scshape]
\item [\textcolor{red}{Exposure time}] For a two-detection requirement, 90\,s exposures place a tighter constraint of $R < 3409$ Gpc$^{-3}$ yr$^{-1}$, in comparison to $R < 3529$ Gpc$^{-3}$ yr$^{-1}$ for 30\,s exposures, assuming a kilonova population with the same intrinsic parameters as AT2017gfo and a uniform-in-cosine distribution of viewing angles. 
\item [\textcolor{red}{Filters}] Including $i$-band observations improves the number of recovered kilonovae by up to 74\% compared to those recovered in $g$ and $r$ bands only, and constrains the kilonova rate to $R < 2749$ Gpc$^{-3}$yr$^{-1}$ assuming the cosine distribution of viewing angles. They are especially useful $\gtrsim$1 day post-merger.
\item [\textcolor{red}{Cadence}] Scheduling three epochs per night, following a $g$-$r$-$i$ filter sequence, is the most efficient cadence strategy; doing this on a nightly basis, in comparison to with a cadence of 3-5 days, leads to an up to 470\% improvement in kilonovae recovered within a given field.
\end{description}

Amongst the above conclusions, our findings regarding the exposure time strategy are most ZTF-specific, given that we assume the same readout and overhead times, and relationship between exposure time and sensitivity, as for ZTF.

Nevertheless, having created realistic simulations that take into account all of the possible challenges that emerge during the scheduling process, these conclusions (the adoption of the aforementioned high-cadence strategies and shorter exposure times, and the inclusion of redder filters) generally hold true for optical survey facilities. We include factors that cannot easily be taken into account in analytic explorations of the same nature, and as a result provide more realistic predictions on rate constraints and relative improvement between different strategies. For example, the upper limit on the kilonova rate for our simulated 30\,s schedule came out to be 1807 Gpc$^{-3}$ yr$^{-1}$ (assuming all kilonovae possess viewing angles equal to 30$^{\circ}$). Compared to the upper limit of 1775 Gpc$^{-3}$ yr$^{-1}$ extracted from ZTF Phase I observations \citep{andreoni2020constraining}, this is only a $\sim$1.8\% difference, thus verifying that our simulated schedules accurately incorporate the effects of weather and telescope-related contingencies. Although there have been various works that explore the effects of different filter combinations for wide-field surveys \citep[e.g.,][]{Andreoni_2019,Bianco_2019}, they usually emphasize the theoretical gains in implementing certain sequences (e.g., displaying the most significant color change); our work has instead explored the more fundamental question of: \textit{which filter combinations yield the highest probability of detection at sufficient SNR?}. The $g$-$r$-$i$ combination outperforms the other combinations in this regard, despite the fact that $g$-$i$-$i$ and $g$-$i$-$g$ sequences have been shown to encode beneficial information about the source. More generally, our work is closest in nature to \cite{SeBi2019}, which probes the performance of different scheduling strategies in a practical setting, but here, we have instead looked into each component of the survey structure -- and its consequent effects on kilonova detection efficiencies and rate constraints -- individually. The recommendations delineated above have been qualitatively stated and theoretically justified in previous works \citep[e.g.,][]{Scolnic_2017,zhu2021kilonova}, but we have concretely quantified the benefit of implementing such strategies with ZTF by novelly determining how much they would constrain the kilonova rate, which is the essential outcome of all serendipitous kilonova searches performed to date \citep[e.g.,][]{DoKe2017,Yang_2017,andreoni2020constraining}.

Although the inclusion of observations in redder filters has been shown to be of benefit in this study, their cadence could possibly be relaxed from the suggested nightly visits due to the longer lasting lightcurve in such bands; this point would be interesting to explore in future studies. We also want to explicitly enumerate some of the simplifications that are likely to affect the results, albeit at a minor level. Our simulations were optimized over 50\% of the survey time, although given the other survey priorities such as high-cadence surveys of the Galactic plane, it is likely that this program would receive $<50$\% of the survey time. Throughout, we also assumed only two kilonova populations, but kilonovae may evolve faster or slower than the rates we assumed. In the case of slower evolution, the detections would still be possible, although for significantly faster evolution, we may miss some of the transients depending on the cadence adopted. We also do not account for efficiency losses due to the image processing pipeline, such as nuclear transients, or the possibility of false positives due to either instrumental effects such as ghosting or astrophysical sources such as cataclysmic variables. 
In reality, kilonovae detected several days post-peak in redder bands may be more challenging to identify in real-time than projected by our study, due to the relative lack of follow-up imagers with sensitivity in those bands. 

Of course, candidate detection with ZTF is not enough to unambiguously identify kilonovae, as we also require follow-up to characterize and classify these sources.
This study therefore encourages the need for automated follow-up infrastructure; this includes both infrastructure designed for triggering and collating observations based on external skymaps, such as the GROWTH Target of Opportunity (ToO) marshal \citep{CoAh2019} and the GRANDMA (Global Rapid Advanced Network Devoted to the Multi-messenger Addicts) pipeline \citep{Antier:2019pzz}, but also alert stream filtering and marshals such as GROWTH's marshal \citep{Kasliwal2018} and AMPEL \citep{Nordin:2019kxt}. In addition, this emphasizes the trend towards singular interfaces to trigger telescope observations, such as the ``Target and Observation Managers'' (TOMs) being built by Las Cumbres Observatory and others \citep{StBo2018}. 

Our recommendations overall for redder filters and higher cadence would tremendously benefit ongoing efforts by other groups to identify faint, fast, and red transients. Furthermore, this modified strategy will provide us with a much stronger chance of kilonova discovery independent of GWs and GRBs.

\section*{Acknowledgements}

We thank Brad Cenko, Mansi Kasliwal, and David Kaplan for the valuable suggestions provided throughout the process of writing this paper. We would also like to thank the anonymous reviewer for their insightful comments. The group at the American University of Sharjah acknowledges a research grant from the Mohammed Bin Rashid Space Centre (UAE), which supported this work. S.~Anand acknowledges support from the GROWTH-PIRE grant 1545949. M.~W.~Coughlin acknowledges support from the National Science Foundation with grant number PHY-2010970. M.B. acknowledges support from the Swedish Research Council (Reg. no. 2020-03330).

Based on observations obtained with the Samuel Oschin Telescope 48-inch and the 60-inch Telescope at the Palomar Observatory as part of the Zwicky Transient Facility project. ZTF is supported by the National Science Foundation under Grant No. AST-1440341 and a collaboration including Caltech, IPAC, the Weizmann Institute for Science, the Oskar Klein Center at Stockholm University, the University of Maryland, the University of Washington (UW), Deutsches Elektronen-Synchrotron and Humboldt University, Los Alamos National Laboratories, the TANGO Consortium of Taiwan, the University of Wisconsin at Milwaukee, and Lawrence Berkeley National Laboratories. Operations are conducted by Caltech Optical Observatories, IPAC, and UW.

\section*{Data Availability}

The data underlying this article will be shared on reasonable request to the corresponding author.



\bibliographystyle{mnras}
\bibliography{references} 








\bsp	
\label{lastpage}
\end{document}